\documentclass[slac_one]{revtex4}
\usepackage{graphicx}
\usepackage{fancyhdr}
\begin{document}

\title{Missing ET Performance in ATLAS} 

%
\author{M.Hodgkinson on behalf of the ATLAS collaboration}
\affiliation{University of Sheffield, Sheffield, UK}

\begin{abstract}
The observation and measurement of missing transverse energy in an event is
a key signature in SUSY searches. This paper describes the reconstruction
and calibration of missing transverse energy in ATLAS, and addresses the
quality of the measurement in a variety of simulated processes which
contain physical missing transverse energy such as $Z$ $\rightarrow$ $\tau^{+}\tau^{-}$ , $A^{0}$/$H^{0}$
$\rightarrow$ $\tau^{+}\tau^{-}$, $t$$\bar{t}$, and SUSY as well as processes which contain no
missing transverse energy such as minimum bias and $Z$+jets. Expectations for
the use of early data in evaluation of the measurement of missing transverse
energy are presented.
\end{abstract}

\maketitle

\thispagestyle{fancy}


\vspace{-0.5in}
\section{INTRODUCTION} 
The ATLAS combined reconstruction software uses several approaches to calculating the missing transverse energy (Missing
ET \cite{csc-ref}). The simplest approach is to use as input all calorimeter cells, which are calibrated based on the cell energy density and position,
 and muons which leave little energy in the calorimeter. This approach is expected to be robust in early data and is constructed from three terms:
The \emph{calo term} is calculated from all calorimeter cells, the \emph{muon term} from muons measured in the muon system
with $|{\eta}|$ $<$ 2.7 and the \emph{cryostat term} corrects for energy deposited in the cryostat between the electromagnetic
and hadronic calorimeters. A more sophisticated approach, called the Refined Missing ET, is to modify these calorimeter cell 
weights according to the object (electron, jet etc) that they are attached to. 

\vspace{-0.2in}
\section{PERFORMANCE}

The linearity is defined to be:
\begin{equation}
\frac{|E_{TMiss}^{True}|-|E_{TMiss}|}{|E_{TMiss}^{True}|}
\end{equation}
where $E_{TMiss}^{True}$ is the true Missing ET calculated from non-interacting truth particles. The performance is shown for a number of samples in Figure 1. In the left plot points with true Missing ET of 20 GeV are from $Z$ $\rightarrow$ $\tau^{+}\tau^{-}$, at 35 GeV from $W$ $\rightarrow$ e/$\mu\nu$, at 68 GeV from semileptonic $t$$\bar{t}$ events, at 124 GeV from $A^{0}$ $\rightarrow$ $\tau^{+}\tau^{-}$ with $m_{A}$ = 800 GeV and at 280 GeV from events with supersymmetric particles with a mass scale of 1 TeV. The right plot shows a more detailed version for $A^{0}$ $\rightarrow$ $\tau^{+}\tau^{-}$ with $m_{A}$ = 800 GeV - the non-linearity at low true Missing ET is due to the finite detector resolution and is not a bias in the algorithm.
\begin{figure}[h]
\begin{center}
$\begin{array}{c@{\hspace{0.5in}}c}
\multicolumn{1}{l}{\mbox{\bf (a)}} &
        \multicolumn{1}{l}{\mbox{\bf (b)}} \\ [-0.53cm]
\hspace{-0.4in}
\includegraphics[width=7.cm,height=3.7cm]{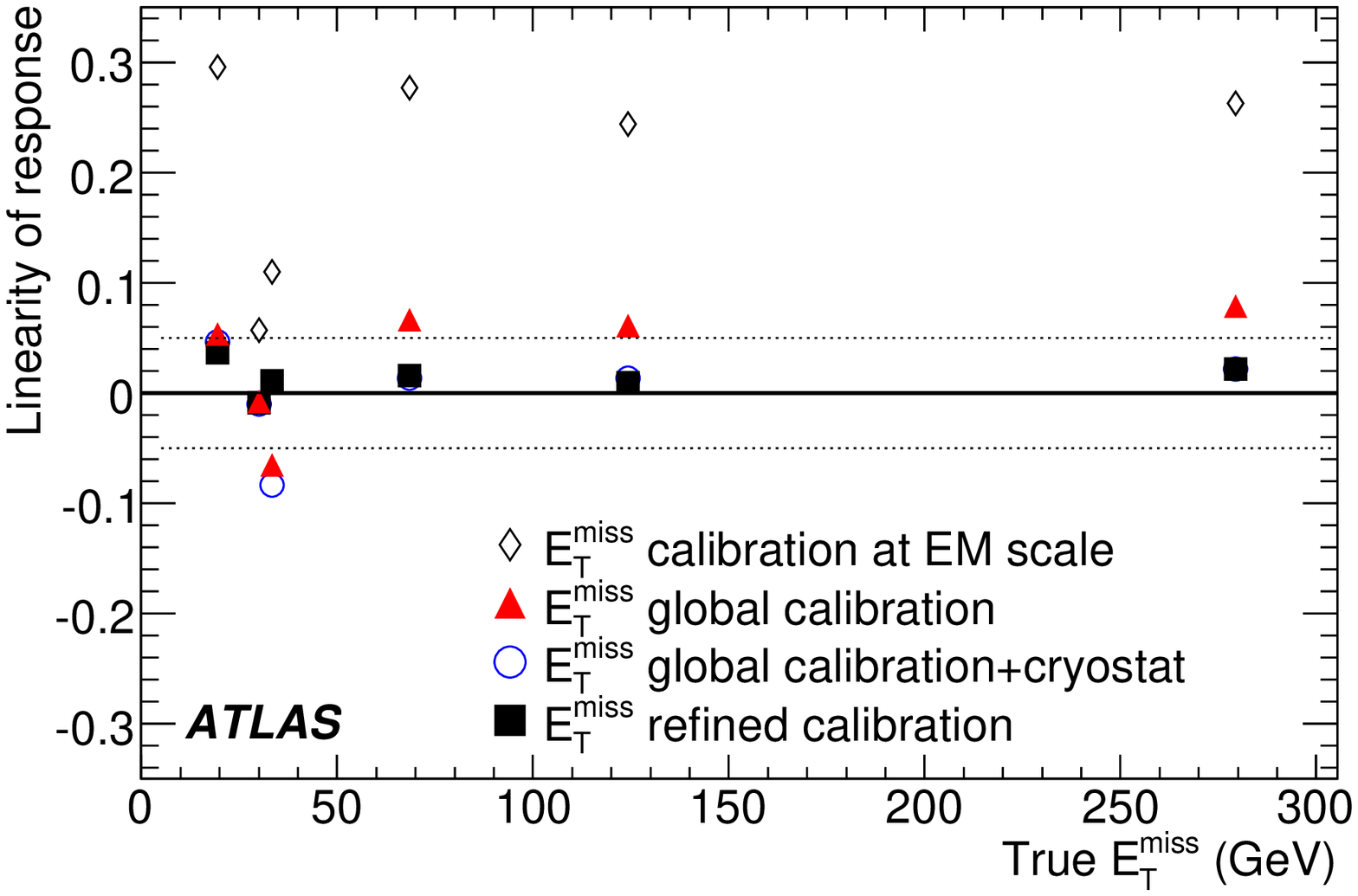} &
        \includegraphics[width=7.cm,height=3.7cm]{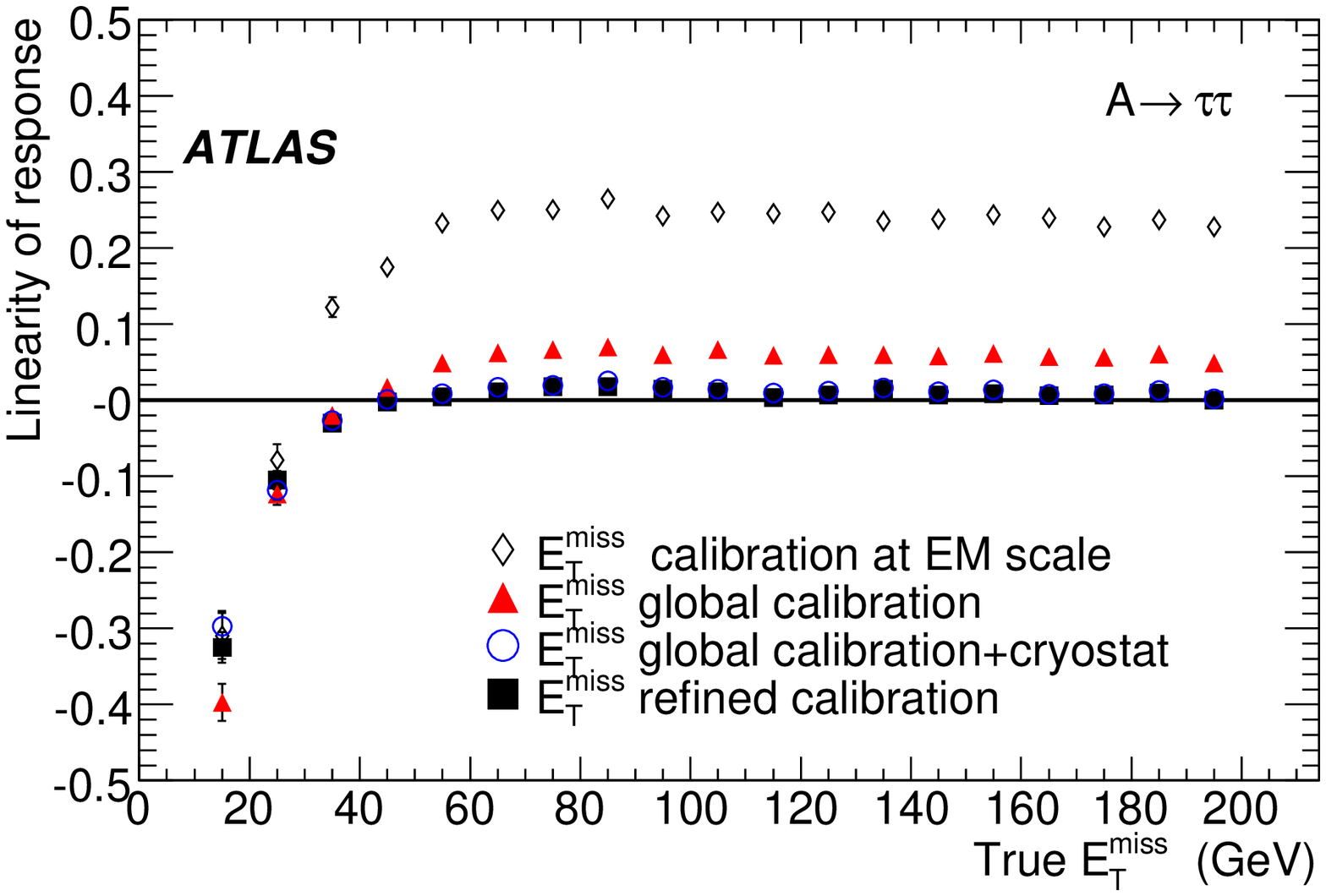} \\ [-0.4cm]
\end{array}$
\end{center}
\caption{Linearity as function of true Missing ET for different stages in the calibration. The global calibration refers to the hadronic calibration cell weights applied to the cells in the \emph{calo term}. }
\label{ETMiss_Lin}
\end{figure}

The resolution is shown as a function of scalar missing transverse energy sum ($\sum{E_{T}}$) for a number of processes in Figure 2. The resolution follows a stochastic behaviour, $a\sqrt{\sum{E_T}}$, with $a$ between 0.53 and 0.57.
\begin{figure}[h]
\begin{center}
$\begin{array}{c@{\hspace{0.5in}}c}
\multicolumn{1}{l}{\mbox{\bf (a)}} &
        \multicolumn{1}{l}{\mbox{\bf (b)}} \\ [-0.53cm]
\hspace{-0.4in}
\includegraphics[width=7.cm,height=3.7cm]{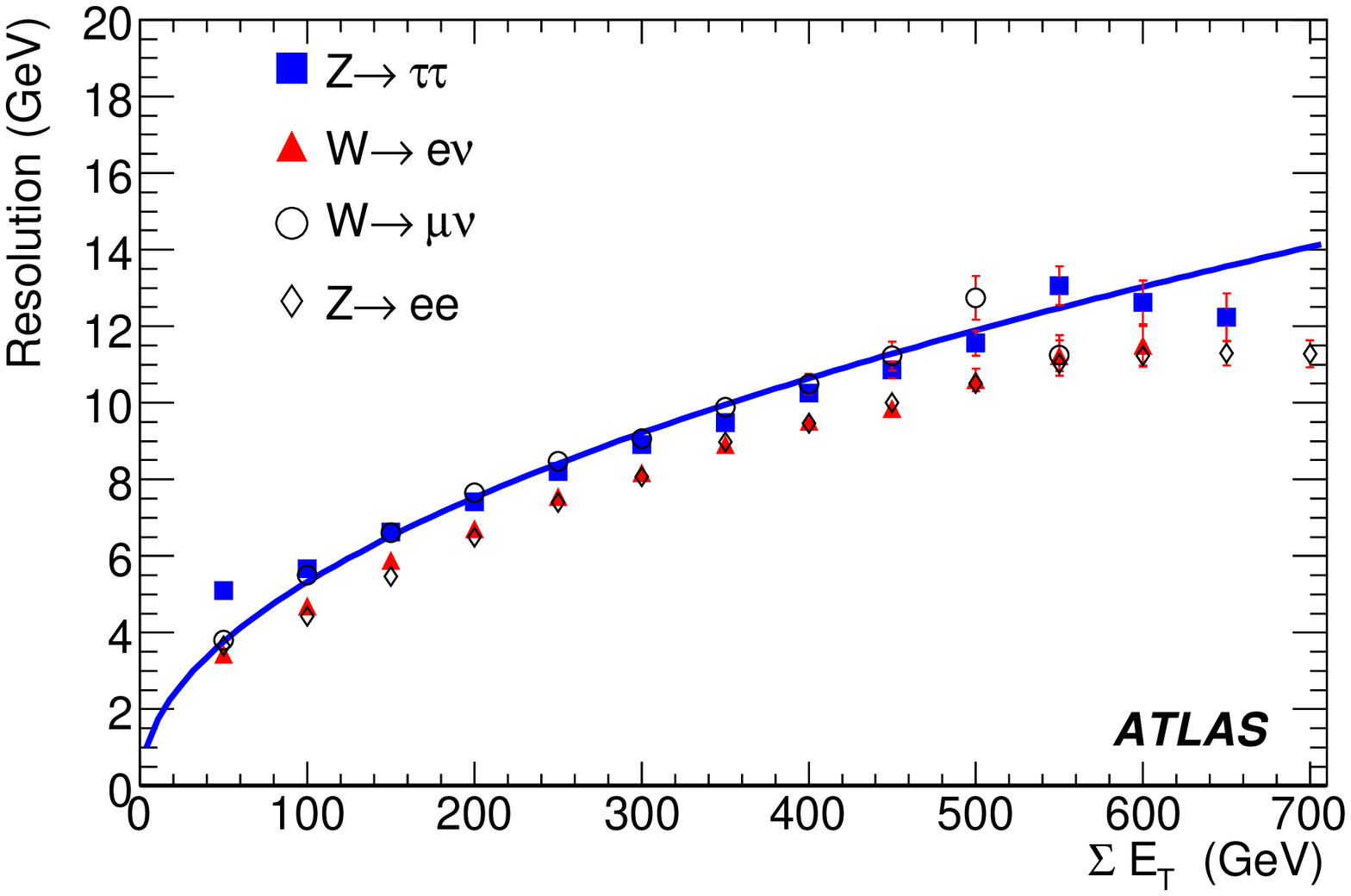} &
        \includegraphics[width=7.cm,height=3.7cm]{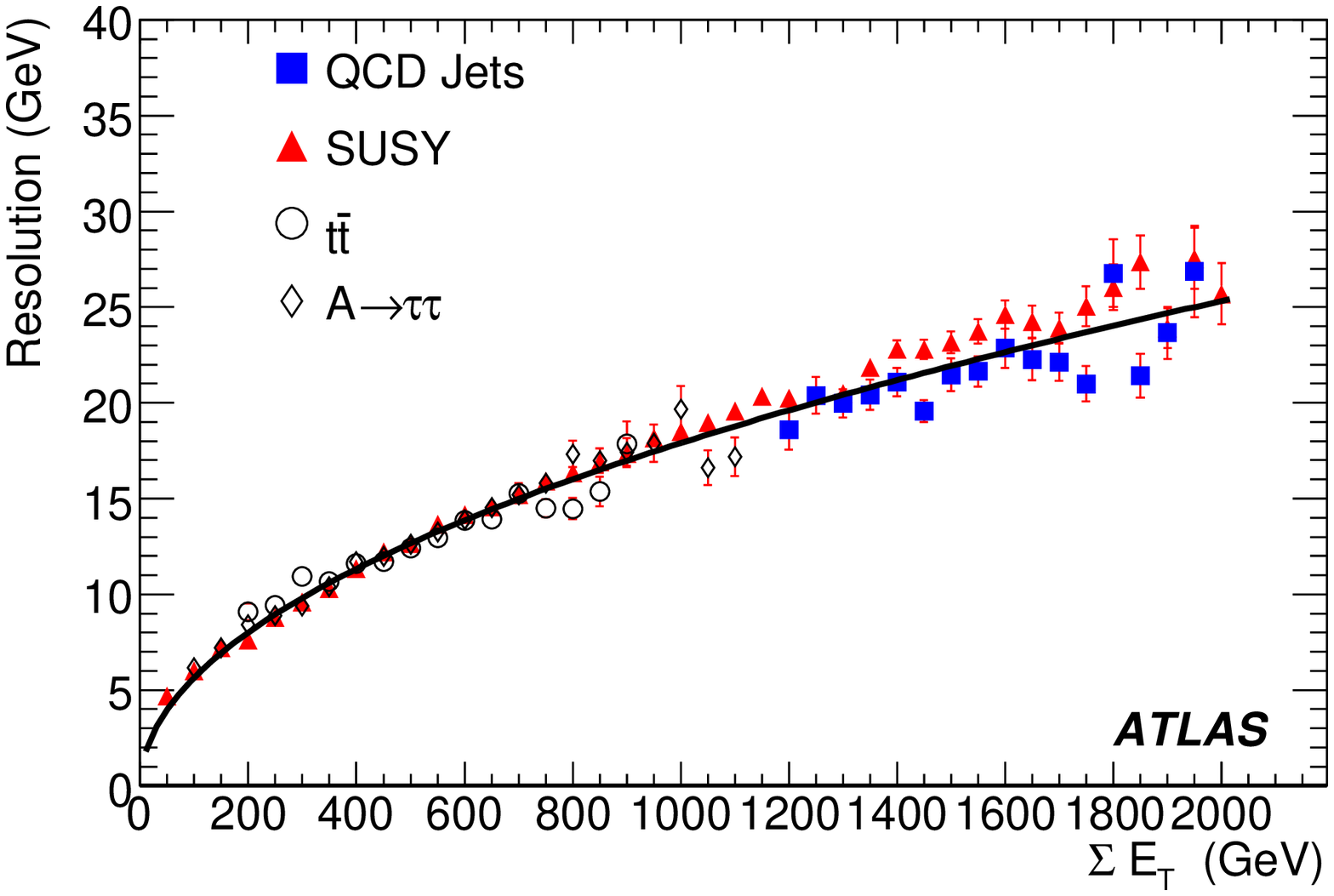} \\ [-0.4cm]
\end{array}$
\end{center}
\caption{Resolution as function  scalar missing transverse energy ($\sum{E_{T}}$) of for a number of physics processes. }
\label{ETMiss_Lin}
\end{figure}
\vspace{-0.3in}
\section{FAKE MISSING ET}
There are a number of sources of fake Missing ET: instrumental effects which can be understood using real data - beam gas, beam halo and noisy or dead cells, inefficiencies in reconstructing high $p_{T}$ muons, fake muons from e.g. jets punching through calorimeters and jet leakage from calorimeters. The effects of fake sources of Missing ET is shown for QCD di-jet events in the left plot of Figure 3 where there is a significant shape difference between the true Missing ET distribution and the fake Missing ET Distribution.  

\begin{figure}[h]
\begin{center}
$\begin{array}{c@{\hspace{0.5in}}c}
\multicolumn{1}{l}{\mbox{\bf (a)}} &
        \multicolumn{1}{l}{\mbox{\bf (b)}} \\ [-0.53cm]
\hspace{-0.2in}
\includegraphics[width=7.cm,height=3.7cm]{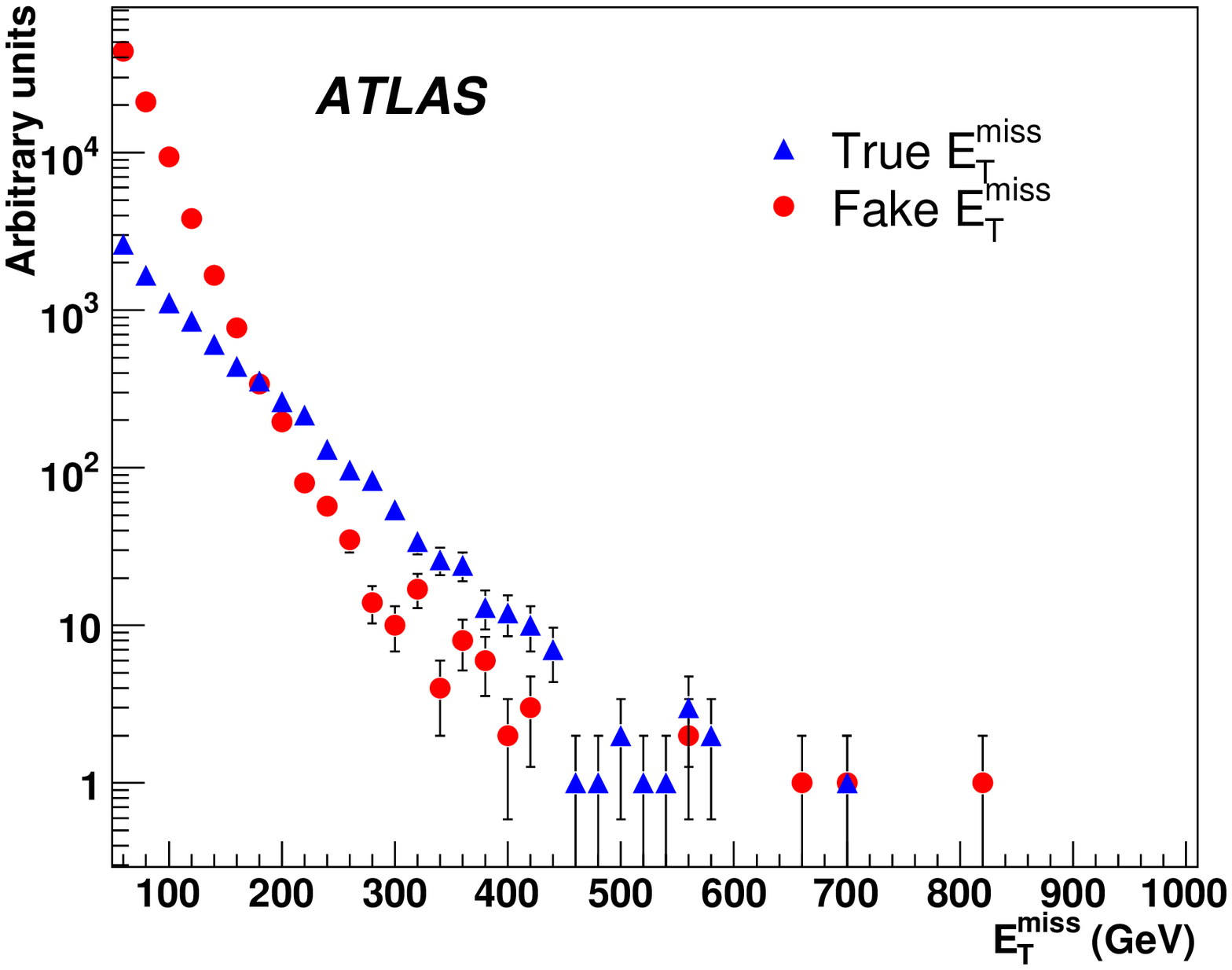} &
        \includegraphics[width=7.cm,height=3.7cm]{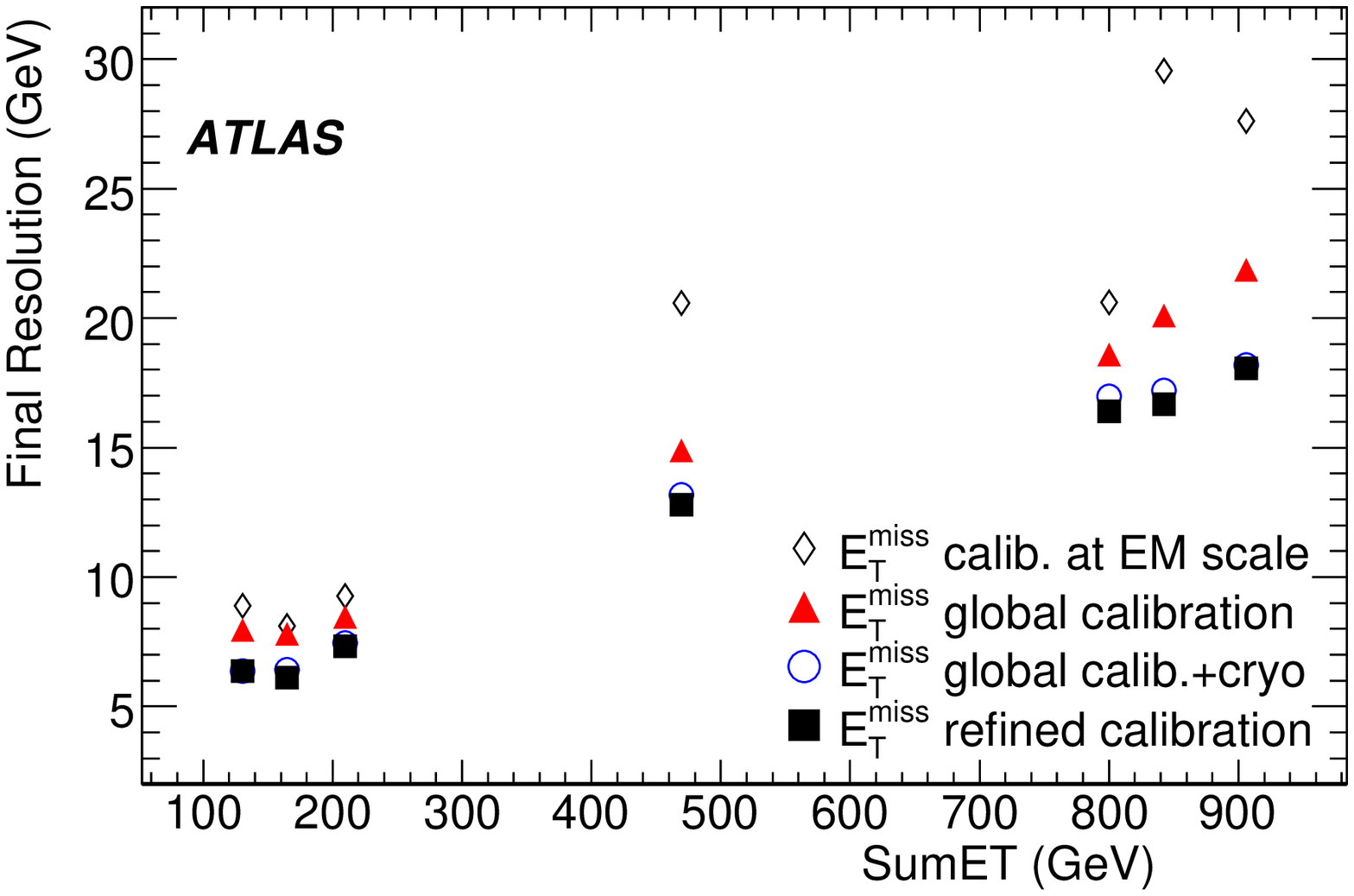} \\ [-0.4cm]
\end{array}$
\end{center}
\caption{True and fake Missing ET distributions for di-jets sample with $p_{T}$ of the hard scatter between 560 and 1120 GeV (left). Missing ET resolution is shown as a function of $\sum{E_{T}}$. $W$ $\rightarrow$ $\mu\nu$ corresponds to $\sum{E_T}$ of 150 GeV, $W$ $\rightarrow$ $e\nu$ to 165 GeV, $Z$ $\rightarrow$ $\tau^{+}\tau^{-}$ to 210 GeV, $t$$\bar{t}$ to 470 GeV, $A^{0}/H^{0}$ $\rightarrow$ $\tau^{+}\tau^{-}$ to 843 GeV, di-jets with $p_{T}$ of the hard scatter in the range 280 to 560 GeV corresponds to 800 GeV and SUSY to 906 GeV (right). }
\label{ETMiss_Lin}
\end{figure}
\vspace{-0.3in}
\section{VALIDATION WITH FIRST DATA}
The first data will be used to calibrate the calorimeter and understand instrumentation failures. Subsequently the $Z$ $\rightarrow$ $\tau^{+}\tau^{-}$ process can be used to determine the Missing ET scale in situ with a systematic uncertainty of 8$\%$. The $W$ $\rightarrow$ $l\nu$ decays can be used to test the Missing ET reconstruction over the energy range 20 to 150 GeV. Top pair production will allow the reconstruction of genuine Missing ET to be tested in an enviroment relevant to Supersummetry (SUSY). The Missing ET resolution is shown as a function of $\sum{E_{T}}$ for various processes at four different stages of the Refined Missing ET calibration in the right plot of Figure 3.

The minimum bias sample will have large statistics in the first data and thus it will be very useful for commissioning the Missing ET reconstruction. Events are required to pass one of three minimum bias triggers, to have at least 20 semiconductor tracker space points to reject empty events and to have at least one good reconstructed track to reject beam gas and halo events. The mean true Missing ET is 0.06 GeV and mean expected Missing ET is 4.3 GeV because of the calorimeter energy resolution and acceptance. At low $\sum{E_{T}}$ the stochastic term dominates the resolution, illustrated in the left plot of Figure 4. The Missing ET is observed to scale with the $\sum{E_{T}}$ as expected when compared to QCD di-jet events (right plot in Figure 4).
\begin{figure}[h]
\begin{center}
$\begin{array}{c@{\hspace{0.5in}}c}
\multicolumn{1}{l}{\mbox{\bf (a)}} &
        \multicolumn{1}{l}{\mbox{\bf (b)}} \\ [-0.53cm]
\hspace{-0.4in}
\includegraphics[width=7.cm,height=3.4cm]{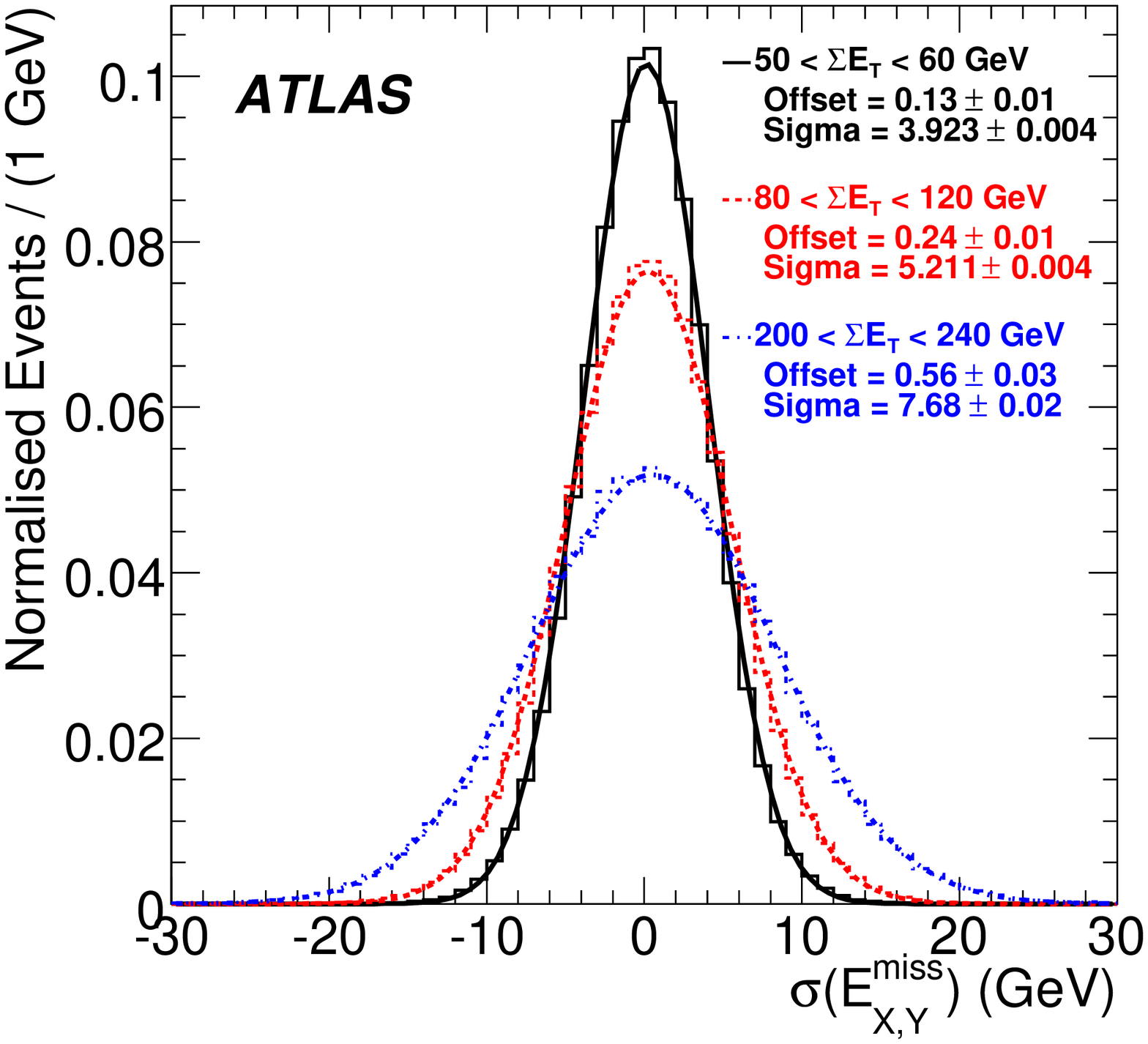} &
        \includegraphics[width=7.cm,height=3.4cm]{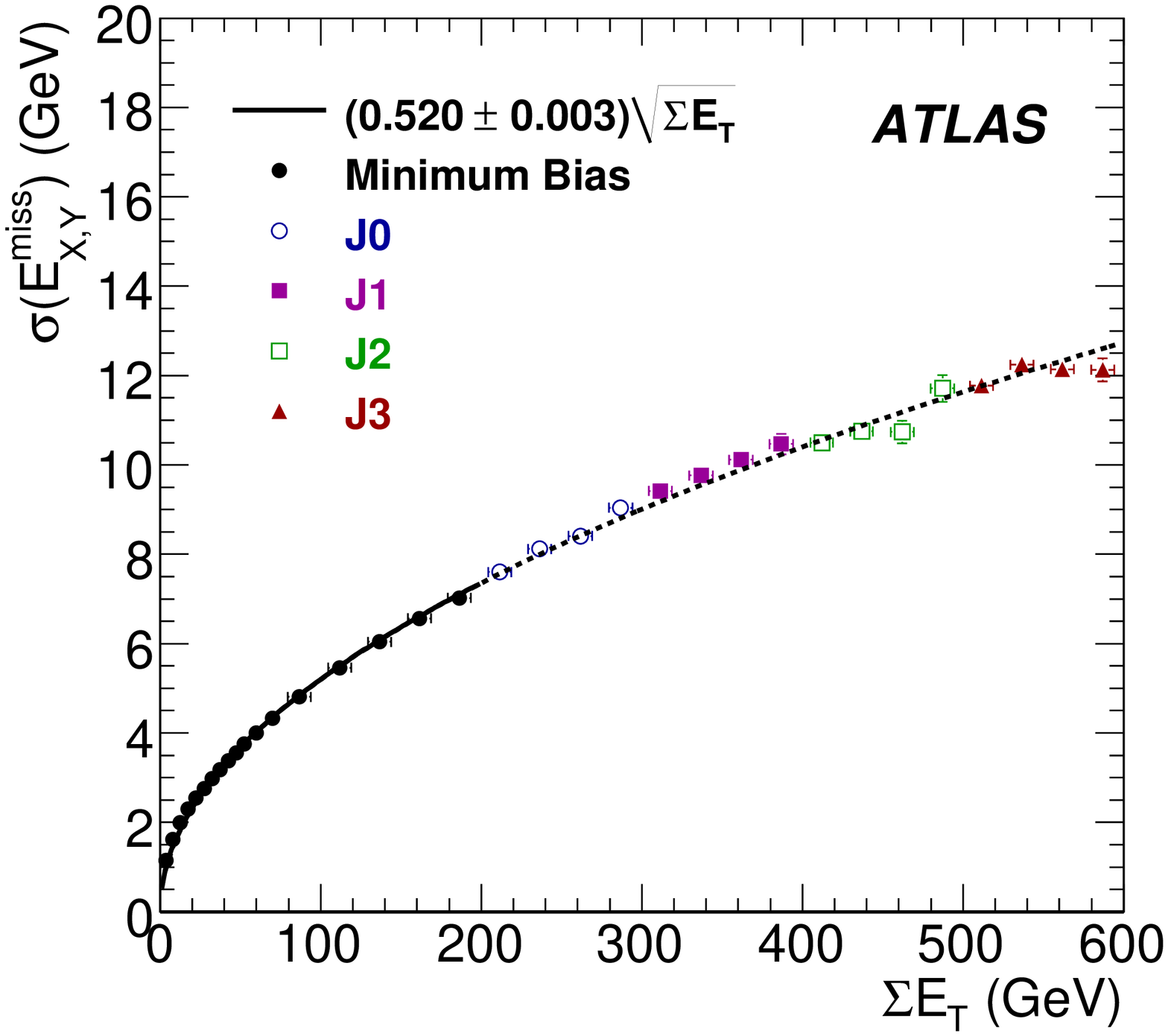} \\ [-0.4cm]
\end{array}$
\end{center}
\caption{Resolution for different $\sum{E_{T}}$ regimes in minimum bias events (left) and Missing ET resolution as function of $\sum{E_{T}}$ for both minimum bias and di-jet events. }
\label{ETMiss_MinBias}
\end{figure}

To select the $Z$ $\rightarrow$ $\tau^{+}\tau^{-}$ sample the Missing ET is required to be larger than 20 GeV to reject QCD events and the transverse mass, calculated from the Missing ET and lepton (defined to be the leading and isolated lepton, electron or muon, with $P_{T}^{l}$ $>$ 15 GeV and $|{\eta}|$ $<$ 2.5), is required to be less than 50 GeV, to suppress semileptonic $W$ decays. A cut on $\sum{E_{T}}$ $<$ 400 GeV further suppresses QCD events. No $b$-tagged jets are allowed in order to suppress bottom and top quark pair production. At least one tau-jet with $p_{T}$ $>$ 15 GeV, $|{\eta}|$ $<$ 2.5 and a track multiplicity of one or three is required. The azimuthal distance between the isolated lepton and tau-jet is required to be in the range 1 - 2.8 radians to reject badly reconstructed events and further suppress backgrounds. The invariant mass of the $\tau\tau$ system shows good sensitivity to the Missing ET Scale (left plot in Figure 5) and allows the determination of this scale with 8$\%$ accuracy.
In the semileptonic top pair sample 7000 events are selected in 200 $pb^{-1}$ when 3 jets with $p_{T}$ $>$ 40 GeV, one more jet with $p_{T}$ $>$ 20 GeV, Missing ET $>$ 20 GeV and an isolated lepton (electron or muon) with $p_{T}$ are required. The effect of scaling the true Missing ET by 0.8 and 1.2 is shown in the right plot of Figure 5 - Gaussian fits indicate the peak shifts by $\pm$ 7 GeV with a statistical error of 0.5 $\%$.

\begin{figure}[!h]
\begin{center}
$\begin{array}{c@{\hspace{0.5in}}c}
\multicolumn{1}{l}{\mbox{\bf (a)}} &
        \multicolumn{1}{l}{\mbox{\bf (b)}} \\ [-0.53cm]
\hspace{-0.4in}
\includegraphics[width=7.cm,height=3.4cm]{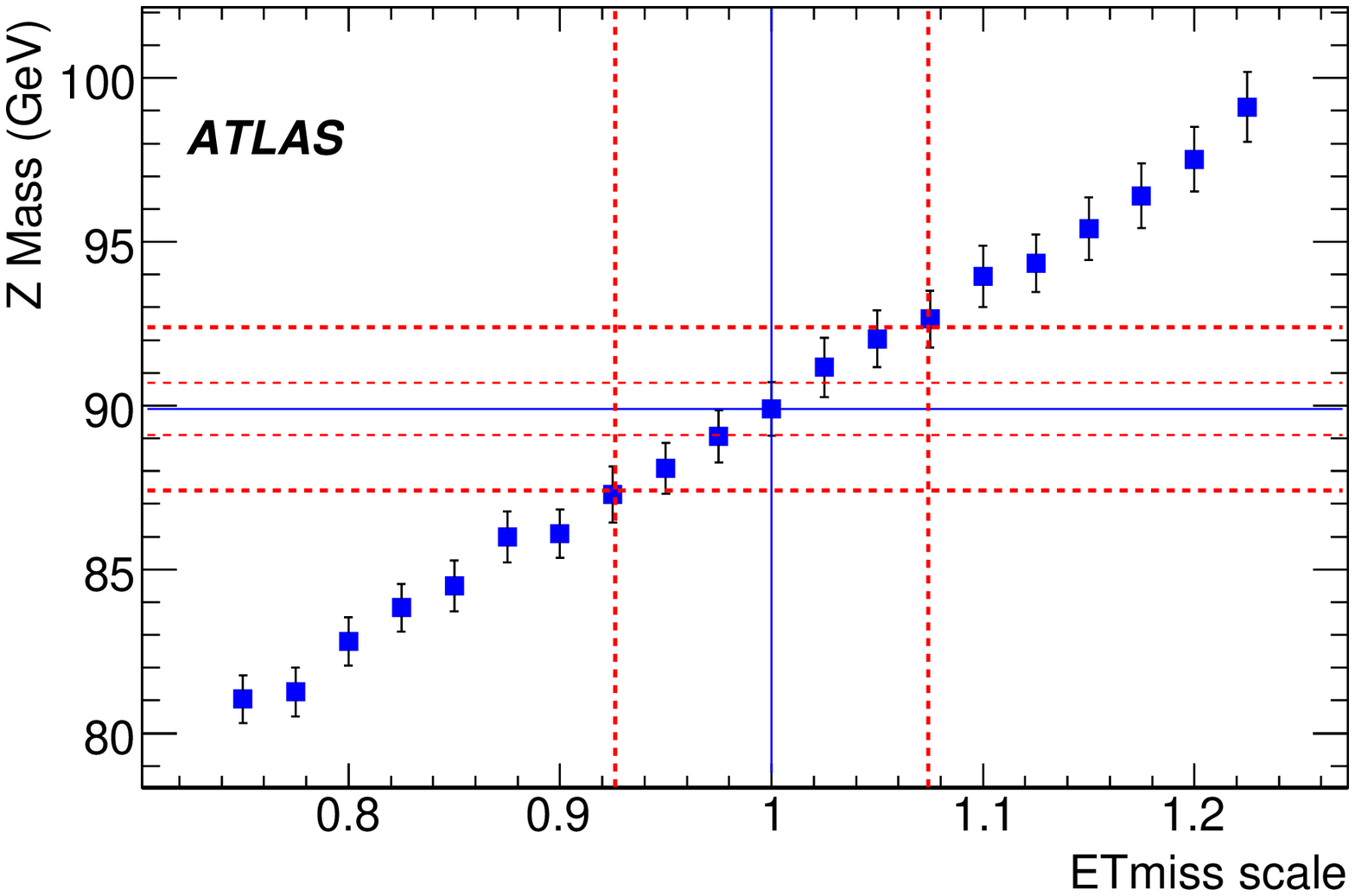} &
        \includegraphics[width=7.cm,height=3.4cm]{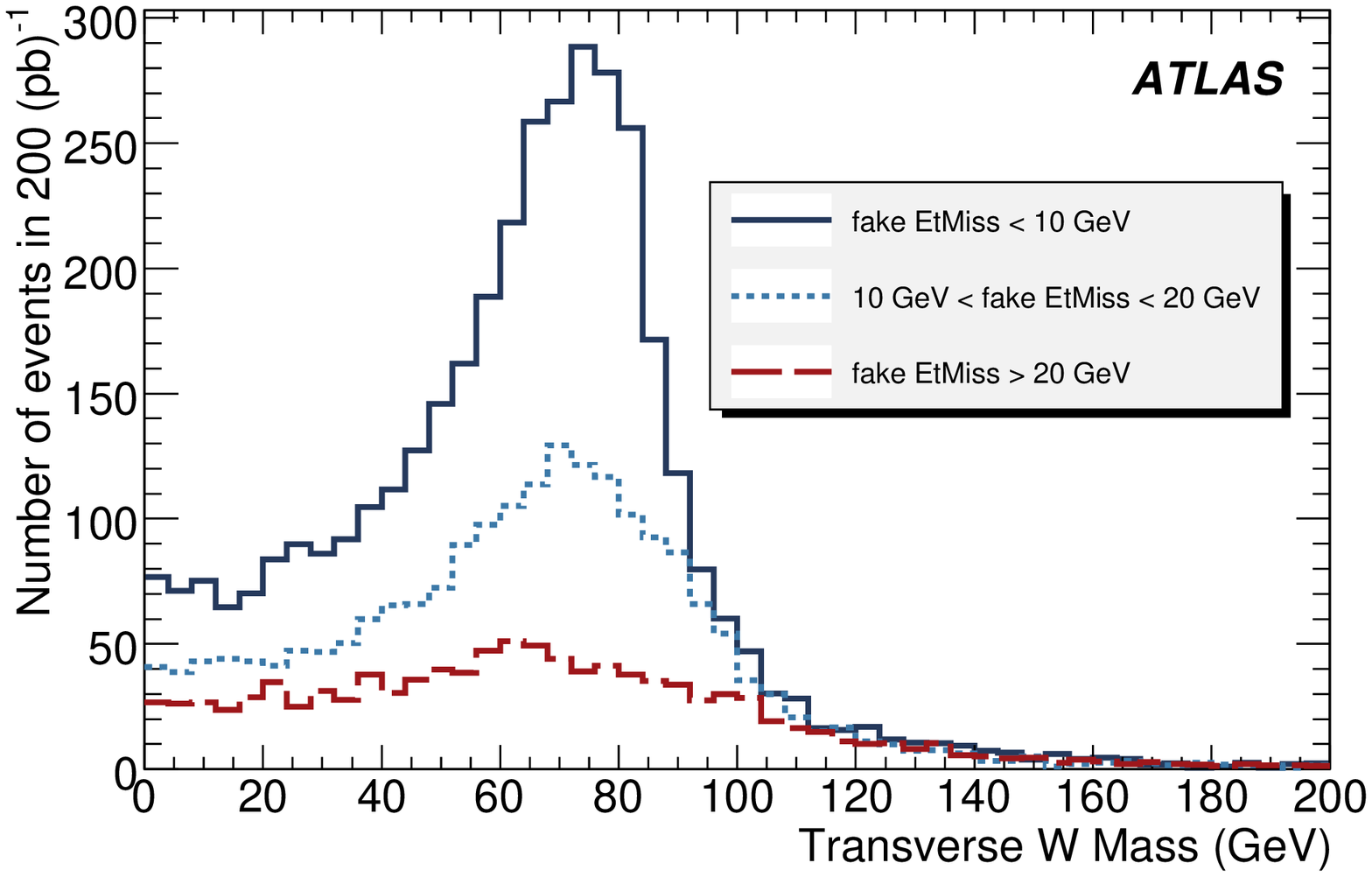} \\ [-0.4cm]
\end{array}$
\end{center}
\caption{The invariant mass of the $\tau\tau$ system as function of the Missing ET scale (left). The transverse mass distribution for three semileptonic top quark pair samples, each with different Missing ET scales (right).}
\label{ETMiss_MinBias}
\end{figure}

\vspace{-0.4in}
\section{CONCLUSIONS}
The linearity of the Missing ET is expected to be within 5$\%$ over a wide range of Missing ET at the beginning of data taking in ATLAS. The resolution follows a stochastic behaviour, $a\sqrt{\sum{E_T}}$, with $a$ between 0.53 and 0.57 except at very low and very high $\sum{E_T}$. Minimum bias events will allow initial validation of the Missing ET reconstruction to be applied. The Missing ET scale can be measured with an 8$\%$ accuracy using $Z$ $\rightarrow$ $\tau^{+}\tau^{-}$ events. The semileptonic top quark pair sample will allow to check for shifts in the Missing ET scale in a kinematic regime relevant for SUSY.

\vspace{-0.2in}

\begin{acknowledgments}
I would like to thank the organisers of the ICHEP08 for the invitation to present this poster and the ATLAS collaboration.
\end{acknowledgments}

\end{document}